# The Coexistence of van Hove Singularities and Superlattice Dirac Points in a Slightly Twisted Graphene Bilayer

Zhao-Dong Chu[§], Wen-Yu He[§], and Lin He*

Department of Physics, Beijing Normal University, Beijing, 100875, People's Republic of China

We consider the electronic structure of a slightly twisted graphene bilayer and show the coexistence of van Hove singularities (VHSs) and superlattice Dirac points in a continuum approximation. The graphene-on-graphene moiré pattern gives rise to a periodic electronic potential, which leads to the emergence of the superlattice Dirac points due to the chiral nature of the charge carriers. Owning to the distinguishing real and reciprocal structures, the sublattice exchange even and odd structures of the twisted graphene bilayer (the two types of commensurate structures) result in two different structures of the superlattice Dirac points. We further calculate the effect of a strain on the low-energy electronic structure of the twisted graphene bilayer and demonstrate that the strain affects the position of the VHSs dramatically.

The charge carriers in graphene possess an internal degree of freedom, which is similar to the chirality for ultrarelativistic particles. Therefore, graphene, the first example of truly two-dimensional crystals [1-3], provides a bridge between condensed matter and high-energy physics [4-7]. The chiral nature of the charge carriers is of central importance to many of graphene's unique electronic properties [4-11]. Probably the most famous example among them is the chiral tunnelling. It was demonstrated that a potential barrier shows no reflection for the chiral electrons in graphene incident in the normal direction, which is a manifestation of the Klein paradox [8-11]. If the potential barrier is replaced by a periodic potential, the chirality of the charge carriers is predicted to result in highly anisotropic behaviours of massless Dirac fermions in graphene and generate new Dirac points at energies $E_{SD} = \pm \hbar v_F |\mathbf{G}|/2$ in monolayer graphene superlattice (here $v_F$ is the Fermi velocity of graphene and $\mathbf{G}$ is the reciprocal superlattice vector of the periodic potential) [12-15].

Experimentally, it was demonstrated that superstructures, such as a corrugated graphene and a moiré pattern induced between the top graphene monolayer and the substrate, act as a weak periodic potential, which generates superlattice Dirac points at an energy determined by the period of the potential [16-19]. These seminal experiments open opportunities for superlattice engineering of electronic properties in graphene. Recent experiments show that graphene-on-graphene moiré pattern can be easily observed in a graphene bilayer with a rotation between the two stacked layers [20-28]. The twisted graphene bilayer provides, to some extent, the most facile method to realize the graphene superlattice with adjustable periods. Therefore, a natural and important problem, which should be addressed, is the effect of the graphene-on-graphene moiré on the electronic structure of the twisted graphene bilayer.

In this Letter, we address the electronic structures of a slightly twisted graphene bilayer. The displaced Dirac cones of the twisted bilayer cross and two intersections of the saddle points along the two cones appear in the presence of interlayer coupling [29]. The saddle points result in two low-energy van Hove singularities (VHSs) in the density of states (DOS) [20-22,26-28]. Beside the VHSs, we show the emergence of superlattice Dirac points in the twisted graphene bilayer. When the two graphene layers are not isolated but sense each other, the graphene-on-graphene moiré gives rise to a periodic electronic potential of triangular symmetry. This periodic potential leads to the emergence of the superlattice Dirac points. The two types of commensurate structures, i.e., the sublattice exchange even (SE-even) and odd (SE-odd) structures, of the twisted graphene bilayer result in quite different structures of the superlattice Dirac points because of their distinguishing real and reciprocal structures. We further calculate the effect of a strain on the low-energy electronic structure of the twisted graphene bilayer and demonstrate that the strain affects the position of the VHSs dramatically. These results indicate that the twisted graphene bilayer provides an ideal system to explore experimentally the superlattice Dirac fermions and is an ideal platform for VHSs engineering.

The conditions for the commensurate structures (the moiré pattern) of the twisted graphene bilayer have been considered previously [30,31]. There are two types of commensurate structures, which are refered as the SE-odd and the SE-even structures, in the twisted graphene bilayer [31]. Fig. 1(a) and Fig. 1(d) show the schematic geometry of the SE-odd and SE-even structures respectively. For the SE-odd structure, the coincident atomic sites are only on the A(A′) sublattice at the origin of the primitive cell (here the two sublattices in layer 1 and 2 are denoted by A, B and A′, B′, respectively). For the SE-even structure, the coincident sites are on the A(A′) and B(B′) sublattices at threefold-symmetric positions in the primitive cell [31].

Owning to the rotation, the Brillouin zones of the two layers have different orientations and the Dirac points of the two layers (K and $K_\theta$ for the first layer and the sub-layer respectively) no longer coincide, as shown in Fig. 1(b) and Fig. 1(e). The shift between the corresponding Dirac points in momentum space is $\Delta K = 2K\sin(\theta/2)$, where θ is the twist angle and $K = 4\pi/3a$ with a ~ 0.246

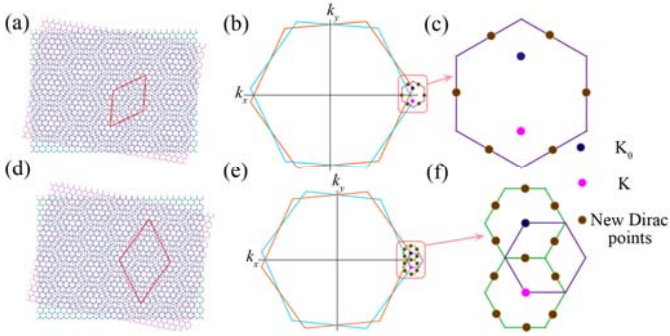
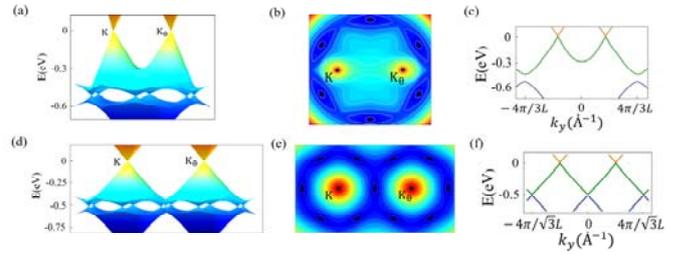

FIG. 1 (color online). (a) Schematic structural model of two misoriented graphene layers with the sublattice exchange odd structure. A periodic moiré pattern is clearly observed. The superlattice of the moiré pattern is plotted. (b) Schematic Brillouin zones of the two graphene layers in panel (a). The twist of the two Brillouin zones and the separation of the two Dirac points, K and $K_\theta$, arises from the rotation between the two graphene layers. (c) Enlarged minizone of a continuum model centered at midpoint between Dirac points K and $K_\theta$. Six superlattice Dirac points are expected to appear at the centre of the minizone boundary. (d) Schematic structural model of two misoriented graphene layers with the sublattice exchange even structure. The superlattice of the moiré pattern is also plotted. (e) Schematic Brillouin zones of the two graphene layers in panel (d). (f) Zoom-in topography of the frame in (e). The purple hexagon is the moiré superlattice Brillouin zone. Green hexagons are minizones of a continuum model with the Dirac points, K and $K_\theta$, at their zone center. The superlattice Dirac points are expected to appear at the centre of the minizones boundary.

FIG. 2 (color online). Energy of charge carriers in a twisted graphene bilayer with (a) the sublattice exchange odd structure and (d) the sublattice exchange even structure. Both of them show the coexistence of the original Dirac points and the new superlattice Dirac points. The graphene-on-graphene moiré generates six new Dirac points in the SE-odd structure and ten superlattice Dirac points in the SE-even structure. In the SE-even structure, the degeneracy of a superlattice Dirac point at the midpoint between the K and $K_\theta$ (from the top view) is lifted due to the interlayer coupling. Density plots of the energy dispersion of the twisted graphene bilayer with (b) the sublattice exchange odd structure and (e) the sublattice exchange even structure. In panel (c) and panel (f) we have a cut of the electronic band structures along $K_\theta$ and K in panel (b) and (e) respectively.

nm. The superlattice Brillouin zones of the SE-odd and SE-even structures are also plotted in Fig. 1(b) and Fig. 1(e) respectively. The relative displacement of the Dirac points ΔK is a reciprocal lattice vector of the moiré superlattice in the SE-even structure but not in the SE-odd structure. This difference in the reciprocal-space is predicted to result in quite different physics in the two types of commensurate structures of the twisted graphene bilayers [29-31]. However, very recent experimental results [20-22,26-28] and theoretical analysis [32] demonstrated that the slightly twisted graphene bilayers share the main physics of the continnum model, i.e., the low-energy linear dispersion and the two low-energy saddle points in the band structure [29].

Although the difference of the SE-odd and the SE-even structures in the reciprocal-space has little effect on the low-energy band structure of the slightly twisted graphene bilayer, we will demonstrate subsequently that it should result in quite different structures of the superlattice Dirac points. Fig. 1(c) and Fig. 1(f) show the first Brillouin zones of a continuum model for the SE-odd and the SE-even structures respectively. When there is interlayer coupling between the two graphene layers, the graphene-on-graphene moiré gives rise to a periodic electronic potential of triangular symmetry. It has been noted that such a periodic potential can create a strong anisotropy in the Fermi velocity around the Dirac point and in the gap opening along the superlattice Brillouin zone boundary [12-15]. The perfect chiral tunneling of the Dirac fermions through a single barrier in the normal direction suggests that the gap should vanish along the direction of the periodic potential, i.e., at the centre of the minizone boundary. The back-scattering of such a graphene superlattice can be neglected because of that the size of the periodic potential is much larger than the nearest-neighbor carbon-carbon distance. The absence of back-scattering is directly related to the formation of the superlattice Dirac points at the centre of the minizone boundary. Therefore, it is expected to generate two different structures of the superlattice Dirac points in the SE-odd and SE-even structures, as shown in Fig. 1(c) and Fig. 1(f).

To further investigate the physics of charge carriers in the graphene superlattice induced by the graphene-on-graphene moiré, we have calculated the electronic band structure of the twisted graphene bilayer in a continuum approximation. For simplicity, we assume that the moiré pattern generates a weak muffin-tin type of periodic potential on the twisted graphene bilayer with the potential value ΔU in a triangular array of disks of diameter $d$ and zero outside of the disks. The spatial period of the superlattice is $L$, which equals to the period of the moiré pattern (here $d < L$ and we demonstrated that the ratio of $d/L$ only has a little effect on the band structure of the twisted graphene bilayer). Then the total Hamiltonian of the twisted graphene bilayer in a periodic potential can be written as

$$H = H_1 + H_2 + H_\perp + \Delta U \sum \overline{(G_\alpha \cdot \vec{k})} I \quad , \qquad (1)$$

where $H_1$ and $H_2$ are the Hamiltonians for each layer, $H_\perp$ is the interaction Hamiltonian between the two layers, $I$ is an identity matrix, and $G_\alpha$ is the potential's reciprocal lattice [12-15,29]. We can use this model to show the new Dirac points induced by the periodic potential for the SE-even and the SE-odd structures (see the supplemental material [33] for details of calculation).

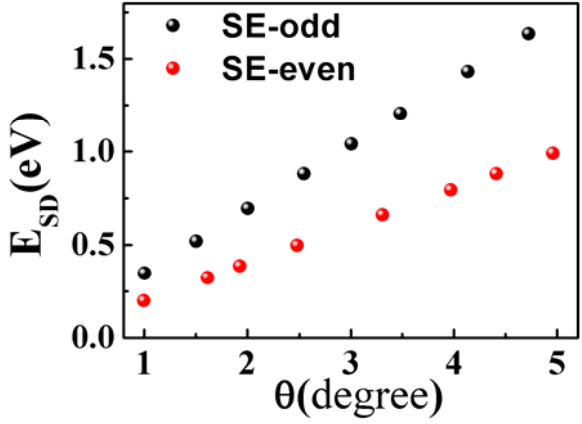

FIG. 3 (color online). The energy of the superlattice Dirac points away from the Dirac point as a function of the rotation angle θ between the two layers. The SE-even and the SE-odd structures show different dependencies due to the difference in their reciprocal-space. Here the rotation angles θ are randomly selected from the two structures.

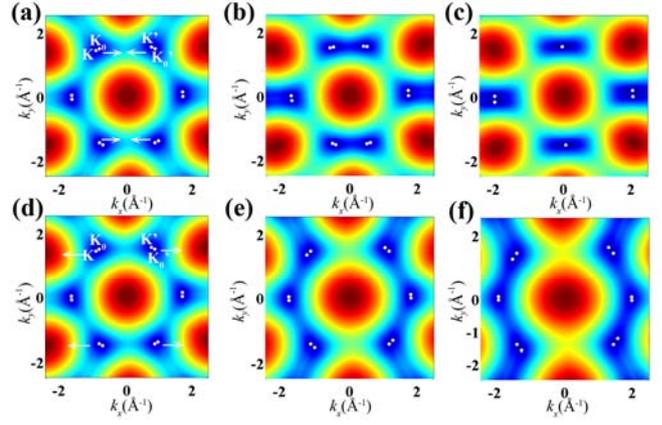

FIG. 4 (color online). The density plots of the energy dispersion of twisted graphene bilayer for (a) {ε = 0, θ = 3.9°}, (b) {ε = 0.14, θ = 3.9°}, and (c) {ε = 0.25, θ = 3.9°} with a tensile strain along the zigzag direction of the top-layer. The density plots of the energy dispersion of twisted graphene bilayer for (d) {ε = 0, θ = 3.9°}, (e) {ε = 0.14, θ = 3.9°}, and (f) {ε = 0.25, θ = 3.9°} with a compress strain along the zigzag direction of the top-layer. The white points are the twelve Dirac points of the twisted graphene bilayer. Four of them, $K$, $K_\theta$, $K'$, and $K'_\theta$, are refered. The arrows show the movement (merging or departure) of the Dirac cones as the strain increases.

Fig. 2(a) and Fig. 2(d) show the energy dispersions of charge carriers in the two types of the commensurate structures respectively. Both of them reveal the coexistence of the two low-energy VHSs and the superlattice Dirac points in the band structure of the twisted graphene bilayer. The gap opening along the superlattice Brillouin zone boundary show strong anisotropy because of the chiral nature of the wave functions. Fig. 2(b) and Fig. 2(e) show the density plots of the energy dispersion of the twisted graphene bilayers. The graphene-on-graphene moiré generates six new Dirac points in the SE-odd structure at the centre of the corresponding superlattice Brillouin zone boundary. In the SE-even structure, the degeneracy of a superlattice Dirac point at the midpoint between the K and $K_\theta$ (from the top view) is lifted due to the interlayer coupling, as shown in Fig. 2(d). Therefore, the graphene-on-graphene moiré generates ten superlattice Dirac points in the SE-even structure. Fig. 2(c) and Fig. 2(f) show the energy dispersions of charge carriers along the K and $K_\theta$ (i.e., at a fixed $k_x$, $k_x$ being zero) in the SE-odd and the SE-even structures, respectively. The result presented in Fig. 2 indicates that the difference of the SE-odd and the SE-even structures in the reciprocal-space really results in quite different structure of the superlattice Dirac points in the twisted graphene bilayer. It also suggests that the twisted graphene bilayers possess much more complex electronic band structures and possible intriguing properties that have not been considered before.

The superlattice Dirac points generated by the graphene-on-graphene moiré appear at energies $E_{SD} = \pm\hbar v_F |\mathbf{G}|/2$, which is identical to that of the single layer graphene superlattice [12-19]. Fig. 3 shows the energy of the superlattice Dirac points away from the Dirac point as a function of the rotation angle θ between the two layers. Owning to the existence of two types of the commensurate structures, the values of $E_{SD}$ do not increase monotone with the rotation angle. However, for a special type of the commensurate structures, either the SE-odd or the SE-even structure, $E_{SD}$ increase with the rotation angle monotonously. The emergence of the superlattice Dirac points, which are symmetrically placed at energies flanking the Dirac point, is manifested by two dips in the DOS. This feature can be easily detected by scanning tunneling spectroscopy (STS) measurements. Here we should point out that experimentally the new massless Dirac fermions in the SE-even structure may be obscured by the states near the saddle points because of that the reciprocal lattice vector of the moiré superlattice equals to the relative displacement of the Dirac points ΔK. However, there exists a large energy window over which the only available states are the new massless Dirac fermions in the SE-odd structure.

Compared with single-layer graphene, twisted graphene bilayers possess two low-energy VHSs. Additionally, the energy of the VHSs in twisted graphene bilayer can be easily tuned by changing the twist angle and the interlayer coupling [20-22,26-29]. These advantages indicate that the twisted graphene bilayer is an ideal platform for VHSs engineering of electronic properties and exploring many attractive phases, such as superconductivity [34]. Below we will demonstrate that a strain also can change the position of the VHSs. For simplicity, we consider the effect of a uniformly tensile strain (or a compressed strain) in the low-energy electronic structure of the twisted graphene bilayer (see Fig. S1 of the supplemental material [33] for the schematic structure of the deformed twisted graphene bilayer). Here we assume that the lattice deformation destroys the chirality of the charge carriers but does not lift the degeneracy of the Dirac points (see the supplemental material [33] for details of calculation). Fig. 4(a)-4(c) shows the energy dispersions of the twisted graphene bilayer with a uniformly tensile strain along the zigzag direction of the

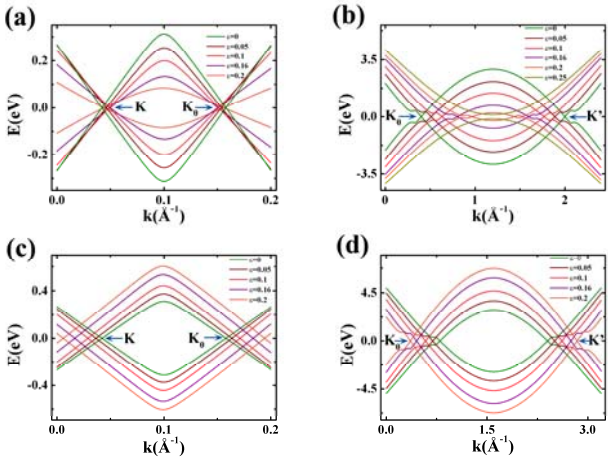

FIG. 4 (color online). (a) A section of the electronic band structures along K and $K_\theta$ for the case of a tensile strain along the zigzag direction of the top layer. The value of $\Delta E_{vhs}$ reduces as the tensile strain increases. (b) A section of the electronic band structures along $K_\theta$ and K' showing the merging of Dirac cones as the tensile strain increases. (c) A cut of the electronic band structures along K and $K_\theta$ for the case of a compress strain along the zigzag direction. The value of $\Delta E_{vhs}$ increases as the strain increases. (d) A section of the electronic band structures along $K_\theta$ and K' showing the departure of Dirac cones as the strain increases.

top-layer. With increasing the lattice deformations and the variation of the nearest-neighbor hopping parameters, the Dirac points move away from the corners K and K'. In small and moderate deformations, the energy difference of the two VHSs decreases as the strain increases, as shown in Figure 5(a). The two Dirac points $K_\theta$ and K' can merge into a single one in a large strain, as shown in Figure 5(b). A compressed strain along the armchair direction of the top-layer leads to similar result of the two VHSs. Fig. 4(d)-4(f) shows the energy dispersions of the twisted graphene bilayer with a uniformly compressed strain along the zigzag direction of the top-layer. The energy difference of the two VHSs increases as the strain increases (a tensile strain along the armchair direction of the top-layer also leads to similar result), as shown in Figure 5(c). The departure of the two Dirac points $K_\theta$ and K' as the strain increase is clear shown in Figure 5(d).

In summary, we consider the graphene-on-graphene moiré on the electronic structure of the twisted graphene bilayer and show the coexistence of VHSs and superlattice Dirac points in such a system. The graphene-on-graphene moiré patterns of the SE-odd and SE-even structures result in two different structures of the superlattice Dirac points. We further calculate the effect of a strain on the low-energy electronic structure of the twisted graphene bilayer and demonstrate that the strain affects the position of the VHSs dramatically. These novel properties presented in this study should be very robust and will be demonstrated sooner rather than later.

During the preparation of the manuscript, we noticed a very recent experimental result that demonstrated the emergence of superlattice Dirac points in the SE-even structure of a twisted graphene bilayer [35]. The obtained band structure by using angle-resolved photoemission spectroscopy is almost identical to that shown in Fig. 2(d) and Fig. 2(e) of this work.

This work was supported by the National Natural Science Foundation of China (Grant No. 11004010), the Fundamental Research Funds for the Central Universities, and the Ministry of Science and Technology of China.

§These authors contributed equally to this paper.
*Email: helin@bnu.edu.cn.

[1] K. S. Novoselov, A. K. Geim, S. V. Morozov, D. Jiang, Y. Zhang, S. V. Dubonos, I. V. Grigorieva, and A. A. Firsov, Science **306**, 666 (2004).
[2] K. S. Novoselov, A. K. Geim, S. V. Morozov, D. Jiang, M. I. Katsnelson, I. V. Grigorieva, S. V. Dubonos, and A. A. Firsov, Nature **438,** 197 (2005).
[3] Y. B. Zhang, Y. W. Tan, H. L. Stormer, and P. Kim, Nature **438** , 201 (2005).
[4] A. H. Castro Neto, F. Guinea, N. M. R. Peres, K. S. Novoselov, and A. K. Geim, Rev. Mod. Phys. **81**, 109 (2009).
[5] S. Das Sarma, S. Adam, E. H. Hwang, and E. Rossi, Rev. Mod. Phys. **83**, 407 (2011).
[6] M. I. Katsnelson and K. S. Novoselov, Solid State Commun. **143**, 3 (2007).
[7] M. A. H. Vozmediano, M. I. Katsnelson, and F. Guinea, Phys. Rep. **496**, 109 (2010).
[8] M. I. Katsnelson, K. S. Novoselov, and A. K. Geim, Nature Phys. **2,** 620 (2006).
[9] Y. Barlas, T. Pereg-Barnea, M. Polini, R. Asgari, and A. H. MacDonald, Phys. Rev. Lett. **98**, 236601 (2007).
[10] I. Brihuega, P. Mallet, C. Bena, S. Bose, C. Michaelis, L. Vitali, F. Varchon, L. Magaud, K. Kern, and J. Y. Veuillen, Phys. Rev. Lett. **101**, 206802 (2008).
[11] A. F. Young and P. Kim, Nature Phys. **5,** 222 (2009).
[12] C.-H. Park, L. Yang, Y.-W. Son, M. L. Cohen, and S. G. Louie, Nature Phys. **4,** 213 (2008).
[13] C.-H. Park, L. Yang, Y.-W. Son, M. L. Cohen, and S. G. Louie, Phys. Rev. Lett. **101**, 126804 (2008).
[14] L. Brey and H. A. Fertig, Phys. Rev. Lett. **103**, 046809 (2009).
[15] M. Killi, S. Wu, and A. Paramekanti, Phys. Rev. Lett. **107**, 086801 (2011).
[16] S. Rusponi, M. Papagno, P. Moras, S. Vlaic, M. Etzkorn, P. M. Sheverdyaeva, D. Pacile, H. Brune, and C. Carbone, Phys. Rev. Lett. **105**, 246803 (2010).
[17] A. L. Vazquez de Parga, F. Calleja, B. Borca, M. C. G. Passeggi, Jr., J. J. Hinarejos, F. Guinea, and R. Miranda, Phys. Rev. Lett. **100**, 056807 (2008).
[18] M. Yankowitz, J. Xue, D. Cormode, J. D. Sanchez-Yamagishi, K. Watanabe, T. Taniguchi, P. Jarillo-Herrero, P. Jacquod, and B. J. LeRoy, Nature Phys. **8,** 382 (2012).
[19] H. Yan, Z.-D. Chu, W. Yan, M. Liu, L. Meng, M. Yang, Y. Fan, J. Wang, R.-F. Dou, Y. Zhang, Z. Liu, J.-C. Nie, and L. He, arXiv: 1209.1689.
[20] G. H. Li, A. Luican, J. M. B. Lopes dos Santos, A. H. Castro Neto, A. Reina, J. Kong and E. Y. Andrei, Nature Phys. **6**, 109 (2010).
[21] A. Luican, G. H. Li, A. Reina, J. Kong, R. R. Nair, K. S. Novoselov, A. K. Geim, and E.Y. Andrei, Phys. Rev. Lett. **106**, 126802 (2011).
[22] W. Yan, M. Liu, R.-F. Dou, L. Meng, Z.-D. Chu, Y. Zhang, Z. Liu, J.-C. Nie, and L. He, Phys. Rev. Lett. **109,** 126801 (2012).
[23] G. M. Rutter, S. Jung, N. N. Klimov, D. B. Newell, N. B. Zhitenev, and J. A. Stroscio, Nature Phys. **7**, 649 (2011).
[24] L. Meng, Y. Zhang, W. Yan, L. Feng, L. He, R.-F. Dou, and J.-C. Nie, Appl. Phys. Lett. **100**, 091601 (2012).
[25] D. L. Miller, K. D. Kubista, G. M. Rutter, M. Ruan, W. A. de Heer, M. Kindermann, P. N. First, and J. A. Stroscio, Nature Phys. **6**, 811 (2010).
[26] L. Meng, Z.-D. Chu, Y. Zhang, J.-Y. Yang, R.-F. Dou, J.-C. Nie, and L. He, Phys. Rev. B **85**, 235453 (2012).

[27] K. Kim, S. Coh, L. Z. Tan, W. Regan, J. M. Yuk, E. Chatterjee, M. F. Crommie, M. L. Cohen, S. G. Louie, and A. Zettl, Phys. Rev. Lett. **108**, 246103 (2012).
[28] I. Brihuega, P. Mallet, H. Gonzalez-Herrero, G. Trambly de Laissardiere, M. M. Ugeda, L. Magaud, J. M. Gomez-Rodriguez, F. Yndurain, and J.-Y. Veuillen, Phys. Rev. Lett. **109**, 196802 (2012)
[29] J. M. B. Lopes dos Santos, N. M. R. Peres, and A. H. Castro Neto, Phys. Rev. Lett. **99**, 256802 (2007).
[30] S. Shallcross, S. Sharma, and O. A. Pankratov, Phys. Rev. Lett. **101**, 056803 (2008).
[31] E. J. Mele, Phys. Rev. B **81**, 161405(R) (2010).
[32] J. M. B. Lopes dos Santos, N. M. R. Peres, and A. H. Castro Neto, Phys. Rev. B **86**, 155449 (2012).
[33] See supplementary material for the detail of analysis and calculation.
[34] R. Nandkishore, L. S. Levitov, and A. V. Chubukov, Nature Phys. **8**, 158 (2012).
[35] T. Ohta, J. T. Robinson, P. J. Feibelman, A. Bostwick, E. Rotenberg, and T. E. Beechem, Phys. Rev. Lett. **109**, 186807 (2012).